\documentclass[conference]{IEEEtran}
\IEEEoverridecommandlockouts

\usepackage{cite}
\usepackage{amsmath,amssymb,amsfonts}
\usepackage{algorithmic}
\usepackage{graphicx}
\usepackage{textcomp}
\usepackage{xcolor}
\usepackage{balance}

 \def\BibTeX{{\rm B\kern-.05em{\sc i\kern-.025em b}\kern-.08em
     T\kern-.1667em\lower.7ex\hbox{E}\kern-.125emX}}
    
\usepackage[T1]{fontenc}
\usepackage{multirow}

\usepackage{url}
\usepackage{subcaption}
\usepackage{flushend}
\usepackage[hidelinks]{hyperref}
\begin{document}

\title{Performance Evaluation of Serverless Edge Computing for Machine Learning Applications}

\author{\IEEEauthorblockN{Quoc Lap Trieu}
\IEEEauthorblockA{
\textit{Western Sydney University}\\
Sydney, Australia\\
0000-0003-0678-374X }
\and
\IEEEauthorblockN{Bahman Javadi}
\IEEEauthorblockA{
\textit{Western Sydney University}\\
Sydney, Australia\\
0000-0003-2351-9801}
\and
\IEEEauthorblockN{Jim Basilakis}
\IEEEauthorblockA{
\textit{Western Sydney University}\\
Sydney, Australia\\
0000-0002-7440-1320 }
\and
\IEEEauthorblockN{Adel N. Toosi}
\IEEEauthorblockA{
\textit{Monash University}\\
Melbourne, Australia \\
0000-0001-5655-5337}
}

\maketitle

\begin{abstract}
Next generation technologies such as smart healthcare, self-driving cars, and
smart cities require new approaches to deal with the network traffic generated by the Internet of Things (IoT) devices, as well as efficient programming models to deploy machine learning techniques. Serverless edge computing is an emerging computing paradigm from the integration of two recent technologies, edge computing and serverless computing, that can possibly address these challenges. However, there is little work to explore the capability and performance of such a technology. In this paper, a comprehensive performance analysis of a serverless edge computing system using popular open-source frameworks, namely, Kubeless,
OpenFaaS, Fission, and funcX is presented. The experiments considered different programming languages, workloads, and the number of concurrent
users. The machine learning workloads have been used to evaluate the performance of the system under different working conditions to provide insights into the best practices. The evaluation results revealed some of the current challenges in serverless edge computing and open research opportunities in this emerging technology for machine learning applications.

\end{abstract}


\section{Introduction}

Machine Learning (ML) applications incorporate data-driven, actionable insights into the user experience and gradually are becoming part of our daily life~\cite{8616889}. From intelligent traffic control and autonomous vehicles to video surveillance and smart healthcare, machine learning applications enable users to more efficiently complete a desired task or action.  What all these applications have in common is the need for advanced data analytics and machine learning models, which are normally hosted in cloud computing infrastructures. The drawback of the cloud is high communication latency between the cloud and the points where data is generated by the Internet of Things (IoT) devices, or where the analysis is needed~\cite{zhang2015cloud,8397448}. To address this challenge and provide a better quality of experience for users, \textit{edge computing} has emerged, in which computing and storage nodes are placed at the network’s edge in close proximity to users~\cite{satyanarayanan2017emergence}. This technology promises to deliver responsive computing services, privacy enforcement and the ability to mask transient cloud outages~\cite{alkhalaileh2020data}. The edge has some distinctive characteristics that makes it more appropriate for applications requiring low latency, mobility support, real-time interactions, online analytics and interplay with the cloud~\cite{shi2016promise}.

Having data analytics and machine learning on the edge rather than in the cloud forces developers to resort to ad hoc solutions specifically tailored to the available infrastructure. The process is heavily manual, task-specific and error-prone, and usually requires good knowledge of the underlying infrastructure~\cite{Mehdipour2019}. Consequently, when faced with large-scale, heterogeneous resource pools, performing effective machine learning is difficult. Thus, there is a need for new programming models and platforms to develop machine learning applications with less operational complexity for the edge. In order to address this challenge in the new edge-cloud computing era, \textit{serverless computing}~\cite{Castro2019} appears to be a promising solution. 

Serverless computing was introduced by prominent cloud providers (companies such as Amazon Web Services (AWS), Microsoft and Google) to have automatic scalability and fine-grained resource billing model for cloud applications while hiding the details of how servers are used and managed. This technology is recognised as a new computing paradigm that allows users to focus on developing business logic and relinquishing them from handling the underlying computing infrastructure~\cite{jonas2019}. The Function-as-a-Service (FaaS) model realises a serverless architecture by allowing users to develop their application as stateless functions to be executed in response to given requests or events~\cite{Castro2019}. What makes the FaaS model particularly attractive is that the underlying hardware and software stack is hidden from cloud users, freeing them from the burden of managing virtual machines, servers and programming frameworks. While serverless architecture is becoming popular in cloud computing~\cite{eismann2021state}, this model has a few inherent constraints that make its adoption in edge computing difficult. Among these limitations, there are short timeouts for function execution, lack of support for stateful applications, poor data management and limits on the amount of computation and memory that a given function can consume~\cite{hellerstein2018serverless}. There are some recent work about performance evaluation of commercial and open-source serverless frameworks, however most of them are cloud-based~\cite{mohanty2018,balla2020open, shahrad2020serverless,216063}. 
In fact, there is a limited number of works on serverless edge computing, the majority of which are focused on conceptual models, architectural challenges and open issues~\cite{Javadi2020,aslanpour2021serverless,9474932}.

%

In this paper, a comprehensive performance analysis of a serverless edge computing system using popular open-source frameworks, namely, Kubeless,\footnote{Kubeless, \url{https://github.com/vmware-archive/kubeless}.} OpenFaaS,\footnote{OpenFaaS, \url{https://github.com/openfaas/faas}.} Fission,\footnote{Fission, \url{https://github.com/fission/fission}} and funcX\footnote{funcX, \url{https://github.com/funcx-faas/funcX}} is presented.
The key contributions of this paper are as follows:
\begin{itemize}
\item Extensive performance evaluation of serverless edge computing's latency considering different programming languages (e.g., Node.js and Python), workloads (e.g., machine learning techniques such as CNN and LSTM), and the number of concurrent users;
\item Investigation of autoscaling behaviour for serverless edge computing under different machine learning workloads to provide insights into the design choices;
\item Analysis of training time and inference time for machine learning workloads for better model selection for serverless edge computing.
\end{itemize}

The rest of the paper is organized as follows. Section~\ref{work} reviews the related work. Section~\ref{system} describes the considered serverless frameworks and the system implementation.
Section~\ref{evaluation} evaluates the performance of the selected frameworks for different workloads.
Finally, Section~\ref{conclusion} provides concluding remarks as well as directions for future work.

\begin{table*}[!t]
\caption{Characterisations of the considered serverless frameworks}
\centering
\begin{tabular}{|p{2.5cm}|p{3cm}|p{3cm}|p{3cm}|p{3cm}|} 
\hline
 \textbf{Features} & \textbf{OpenFaaS} & \textbf{Kubeless} & \textbf{Fission} & \textbf{funcX} \\ 
\hline
\textbf{Supported Languages } & Python, C\#, Go, Node.js, Ruby  and  custom  containers & Python,   Node.js,   Ruby, PHP, Go, Java, .NET and custom containers &  Python,   Node.js,   Ruby,Perl,   Go,   Bash,   .NET, PHP and   custom   containers &  Python \\
\hline
\textbf{Container orchestration} & Kubernetes, Docker Swarm  &  Kubernetes & Kubernetes & Singularity,  Shifter, Docker\\
\hline
\textbf{Auto scaling metric} & CPU utilization,  QPS  and  custom metrics & CPU utilization, QPS and custom metrics & CPU utilization & QPS \\
\hline
\textbf{Triggers} & HTTP, event, schedule & HTTP, event, schedule& HTTP, event, schedule  & HTTP, GlobusAutomate \\
\hline
\textbf{Maximum Walltime} & NA & NA& Undefined & No limit   \\
\hline
\textbf{Message queue} & NATS, Kafka & NATS, Kafka & NATS, Azure storage queue & Redis   \\
\hline
\textbf{Monitoring  tool} & Prometheus & Prometheus & Istio & funcX agent \\
\hline
\textbf{State} & Stateless & Stateless & Stateless & State-full\\
\hline
\textbf{Development Language} & Go & Go & Go & Python\\
\hline
\textbf{Licences} & MIT & Apache 2.0 & Apache 2.0 & Apache 2.0\\
\hline

\end{tabular}
\vspace{0.3cm}

\label{tab:frameworks}
\end{table*}

\section{Related work}
\label{work}
Serverless computing and its popular subdivision of Function-as-a-Service (FaaS) are relatively new technologies that build up the next phase of cloud computing~\cite{Schleier-Smith2021}. In serverless computing, programmers create applications using high-level abstractions offered by the cloud provider. FaaS is the core enabler of serverless~\cite{jonas2019} which offers an event-driven execution model running the server-side logic in stateless containers. Given the recent efforts to adopt the serverless model in edge computing, it is essential to test and carry out performance analysis of serverless frameworks and tools on the edge to understand better the performance gaps and where to focus improvement efforts.

Lee \textit{et al.}~\cite{lee2018} performed a performance analysis of the serverless computing services from four major cloud providers, i.e., AWS, Azure, Google Cloud and IBM Cloud. Similarly, Kuhlenkamp \textit{et al}.~\cite{Kuhlenkamp2019} propose a new evaluation method and metrics to evaluate serverless big data processing applications on AWS Lambda, Google Cloud Functions, IBM Cloud Function, and Microsoft Azure Functions. Mohanty \textit{et al.}~\cite{mohanty2018} studied multiple open source serverless platforms, including  Fission, Kubeless and OpenFaaS, running on a Kubernetes cluster and comprehensively compared the features and architecture of each platform in order to identify the most suitable platform for production environments. A similar study has been done by Bella \textit{et al.}~\cite{balla2020open} in which they considered both IO intensive and compute intensive workloads. These studies, however, are mainly focused on clouds and do not consider the computational limitations of edge devices such as as Raspberry Pis.

Given the attractive prospects of the serverless at the edge~\cite{aslanpour2021serverless}, attention has been paid to the design and implementation of serverless computing frameworks on the edge environments. Rausch \textit{et al.}~\cite{Rausch2019} propose a serverless platform for building and operating edge AI applications. In another work~\cite{Rausch2021}, they propose a container scheduling system that enables serverless platforms to make efficient use of edge infrastructures. Baresi \textit{et al.}~\cite{baresi2017} discuss the adoption of serverless architectures on edge and proposes a new serverless architecture that can be suitable for an edge environment. Pfandzelter~\textit{et al.}~\cite{Pfandzelter2020} present a lightweight platform for FaaS system on edge environments and compared it to Kubeless and Lean OpenWhisk. These studies focus on the design and implementation of serverless edge computing platforms. However, we focus on the performance evaluation of existing open-source platforms at the edge.

Similar to our work, Palade \textit{et al.}~\cite{palade2019} focus on the performance evaluation of open source platforms, namely Kubeless, OpenFaaS, Knative and Apache OpenWhisk at the edge. While we focus on machine learning applications, their evaluation has been done on functions with minimal overhead in terms of business logic. Tzenetopoulos \textit{et al.}~\cite{Tzenetopoulos2021} evaluate the performance of OpenFaaS, OpenWhisk, and Lean OpenWhisk in hybrid edge-cloud environments. Their work differs from ours as they focus on Optical Character Recognition (OCR) application for workload generation, and they have a mixture of VMs and Raspberry Pis in their setup. Javed \textit{et al.}~\cite{javed2021} analyzed the performance of serverless platforms, including OpenFaaS, AWS Greengrass, and Apache OpenWhisk, on the edge devices and compared it with public cloud serverless offerings such as AWS Lambda and Azure Functions. Different to our work, they performed their analysis using CPU, Memory and disk-intensive applications such as matrix multiplication on a cluster of Raspberry Pi devices. 
Bac \textit{et al.}~\cite{9687209} proposed an architecture for deploying machine learning workload as serverless functions in the edge environment using Knative, however, they did not consider different serverless frameworks.

This paper focuses on more realistic workloads including machine learning applications with well-known models such as CNN and LSTM. In addition, more comprehensive analysis on impact of autoscaling and the behaviour of serverless frameworks for different programming languages have been presented, which are missing in the literature.

\section{System overview}
\label{system}

In this section, we first review our selected open-source serverless frameworks, and then we present the system architecture for evaluation of these serverless frameworks in the edge computing configuration.

\subsection{Open serverless frameworks}

In  this  subsection, we describe  four  open  source  serverless frameworks, namely  OpenFaaS, Kubeless, Fission, and funcX. We chose these frameworks due to their popularity and distinct features as listed in Table~\ref{tab:frameworks}. OpenFaaS, Kubeless and Fission utilize container orchestration to manage the networking and lifecycle of the containers, whereas  funcX  may  be  deployed  with  or  without  container orchestration. We present a brief summary of these frameworks and highlighting their main features.

\textit{OpenFaaS} is one of the well-known  open-source   serverless   frameworks   for Docker  and  Kubernetes.  Developers only need to supply function and  handler and  OpenFaaS  CLI will handle  the  packaging  of  the  function  into  a  Docker container. The container includes a function watchdog, 
which is responsible for processing incoming event triggers and for initializing and monitoring the functional logic of the container. An API  gateway provides  an  external interface to the functions, collects metrics and handles scaling by interacting with the container orchestration plugin. According to the default settings, functions auto-scale up or down depending on the queries
per second (QPS), by utilizing Prometheus\footnote{Prometheus,\url{https://prometheus.io}.} alert manager and monitoring. In addition,
OpenFaaS utilizes a message bus for asynchronous function invocations.

\textit{Kubeless} is a Kubernetes-native serverless framework that utilizes Custom Resource Definitions (CRDs) to extend the Kubernetes API and create functions as custom objects. This enables developers to use the native Kubernetes APIs to interact with the functions. It leverages Kubernetes resources to provide auto-scaling, API routing, monitoring, and troubleshooting. The Kubeless controller watches these custom resources and launches runtimes on-demand. The controller dynamically injects the functions' code into the runtimes and make them available over HTTP or via a PubSub mechanism.

\textit{Fission} is an open-source serverless computing framework built on top of Kubernetes and uses many Kubernetes-native concepts.  The framework executes a function inside
an environment that contains a web server and a dynamic language-specific loader required to run the function [18]. An executor controls how function pods are created and scaled. Pods are the smallest deployable units of computing that you can create and manage in Kubernetes.
One of the features of Fission is that it can be configured to maintain a pool of containers so that requests are served with very low latencies. Fission provides autoscaling for functions based on CPU utilization. Fission integrates with Istio,\footnote{Istio, \url{https://istio.io/}.} which is an open platform to connect, manage and secure micro-services. Also, users get the ability to monitor function usage and trace request latency through dashboards.

\textit{funcX} is the federated serverless framework that enables the execution of functions across heterogeneous, distributed resources. funcX has a logical interface to a computational resource which is called \textit{endpoint}, to allow the funcX service to dispatch function invocations to that resource. Each endpoint comprises of agent, mangers and workers for remote execution of functions in a secure, scalable, and reliable manner with the ability to keep the state of the functions~\cite{chard2019serverless}. Currently, funcX only supports queries per second (QPS) as the metric for autoscaling.
The funcX service provides a REST API for registering functions and endpoints, and for executing functions, managing their execution, and retrieving results. In contrast to other frameworks, funcX relies on AWS-hosted databases, caches, and Web serving infrastructure to reduce operational overhead, elastically scale resources, and provide high availability~\cite{chard2020funcx}.

\subsection{System implementation}
\label{system-implementation}
\begin{figure}[!t]
   \centering
    \includegraphics[width=\columnwidth]{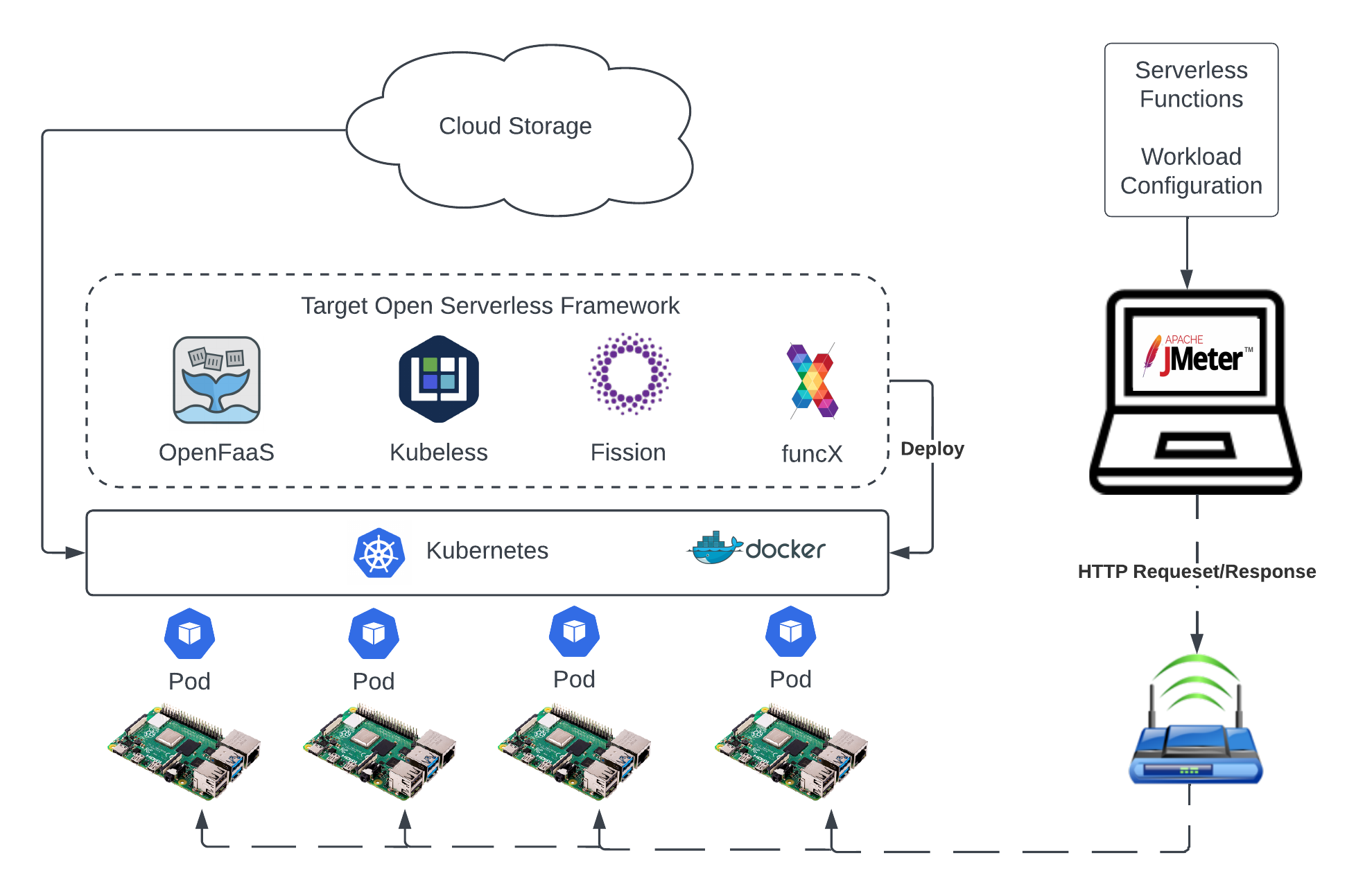}
   \caption{ Serverless edge computing implementation.}
   \label{fig:implementation}
\end{figure}

In order to evaluate the selected frameworks, we set up a serverless edge computing system using a cluster of Raspberry Pis as depicted in Fig~\ref{fig:implementation}. There are four Raspberry Pis 4 Model~B with a 1.5GHz 64-bit quad-core ARM processor and 4GB memeory running Ubuntu 20.04.3 LTS. We also utilize Microk8s\footnote{Microk8s, \url{https://microk8s.io/}.} as the container orchestration, which is a lightweight distribution of Kubernetes built for IoT and edge computing devices. The Kubernetes cluster can also access to a cloud storage (i.e., AWS S3) to download the required files and models for the system workload.

As illustrated in Fig~\ref{fig:implementation}, all four serverless frameworks are deployed separately on the Raspberry Pis cluster which is managed by Kubernetes. The cluster nodes are interconnected using a local WiFi network. For the sake of accuracy, the experiments are run on an isolated network to reduce the impact of network overheads. The system workload and requests are generated by a test machine running Apache JMeter\footnote{JMeter, \url{https://jmeter.apache.org/}.}  v5.4 on the same local network to trigger HTTP requests that invoke functions deployed on each serverless framework. This process is performed through a distributed load testing procedure and orchestrated using the test machine that has a JMeter client installed. The test machine is a laptop running macOS with 2.3GHz 8-core Intel Core i9 and 16GB RAM. The master node of the Kubernetes cluster is also located in the test machine.

\section{Performance evaluation}
\label{evaluation}

In this section, we provide the performance evaluation results and discussions on various serverless frameworks under different workloads.  The main performance metric is \textit{response time}. The response time consists of the time it takes a request to reach the edge device, the time to execute the function on the device, and the time it takes the response to arrive to the JMeter client. Other metrics such as \textit{success rate} and \textit{number of pods} are also reported for some experiments.

\subsection{Experimental setup}
\label{setup}

We configure JMeter to generate HTTP requests that invoke the functions deployed on each framework. The JMeter tool is set up to send 1000 requests with different levels of concurrency ($n=$ 1, 5, 10, 15 and 20). These concurrent requests/users affect the number of simultaneous requests received by the framework and are determined based on the computational power available in the cluster. We also measure the impact of auto-scaling on response time for different workloads. The CPU utilization is used as a metric to perform the auto-scaling, and is set to 50\% for all frameworks in all experiments. When utilization exceeds this threshold, the creation of a new pod is triggered. All frameworks, except funcX adopt the Kubernetes’ Horizontal Pod Autoscaler to perform scaling based on the CPU utilization.

\begin{table*}[htbp]
\caption{Workload configuration}
\centering
\label{workload}
\begin{tabular}{|c|c|c|l|}
\hline
\textbf{Workload}  &  \textbf{Content} & \textbf{Language} & \textbf{Configuration} \\
\hline
Helloworld &  Return request & Python and Node.js & \\
\hline
CNN &  Inference for a CNN model  &  Python & Steps; 128; 1D CNN + Pooling: 4; Dense: 1 \\
\hline
LSTM & Inference for a LSTM model & Python &  Steps: 128; Size: 27; LSTM layers: 2, Dense: 1\\
\hline
\end{tabular}
\end{table*}

We used three different application workloads, including a simple \textit{Helloworld} function as well as two machine learning models based on \textit{Convolutional Neural Networks} (CNN) and \textit{Long short-term memory} (LSTM). These workloads and their configurations are listed in Table~\ref{workload}. 
The Helloworld function written in \textit{Node.js} and \textit{Python} and receives HTTP request and replies with
a confirmation message. This function is used to measure the overhead of each framework for two common programming languages. Each experiment is repeated 3 times in order to maintain statistical accuracy.


CNNs and LSTMs are common deep learning models that have widely been used in image, speech and text recognition, and characteristically use a set of common computations. The CNN has additional sparsely connected convolutional layers compared to its subsequent fully connected dense layers that are optimised for capturing spatial and temporal inputs. In this paper, a one dimensional CNN (Conv1D) is used that specialises in sequences and time-series data. The LSTM consists of a series of gates that unroll programmatically in a sequence that attempts to also model time or sequence-dependent behaviour. The gates within the LSTM cells control information flow within the data sequence, determining whether information is to be remembered or forgotten. 

The machine learning workload consists of classification on human activity recognition tasks using smartphone accelerometer and gyroscope data available from the UCI repository.\footnote{UCI, \url{https://archive.ics.uci.edu/ml/datasets/human+activity+recognition+using+smartphones}.}
There are over 10k instances of data consisting of nine channels of input and six activity tasks to be recognised in the output. The CNN and LSTM models used for training and classification of the time-series data are adopted from the open source Human Activity Recognition (HAR) project in Python.\footnote{HAR, \url{https://github.com/bhimmetoglu/time-series-medicine}.} The CNN model is trained using 1000 epochs in batches of 600, while LSTM model training requires only 15 epochs that is unbatched. Training has been conducted on the test machine (x86) and repeated 3 times resulting in the training times and accuracy shown in Table~\ref{training}. On completion of training, the model weights are uploaded to the cloud storage as illustrated in Fig.~\ref{fig:implementation}, which are subsequently downloaded only once by edge devices to execute the workloads on either CNN or LSTM activity classification. A remote procedure call (RPC) from a client program running on JMeter is used to simulate sending of an input stream of human activity data to the machine learning algorithms deployed on the framework, that will respond with an output of the classified activity\footnote{Workload, \url{https://github.com/SDC-Lab/ServerlessWorklaod}}.

\subsection{Impact of concurrent users}
In this experiment, we measure the response time of Helloworld function for different levels of concurrency in both Python and Node.js for all serverless frameworks. Since funcX does not support Node.js, we only report results of the Python function for this framework. We used
Helloworld function to have minimal overhead in terms of the function
logic and its dependencies. We deploy the function on each
framework and invoke it through HTTP. We disable autoscaling and run a single pod per edge device with a fixed number of concurrency in each experiment. By doing so, we avoid possible impact of autoscaling on response times when scaling in/out functions, i.e., when creating new function pods. 

\begin{figure*}
     \centering
     \begin{subfigure}[b]{0.45\textwidth}
         \centering
         \includegraphics[width=\textwidth]{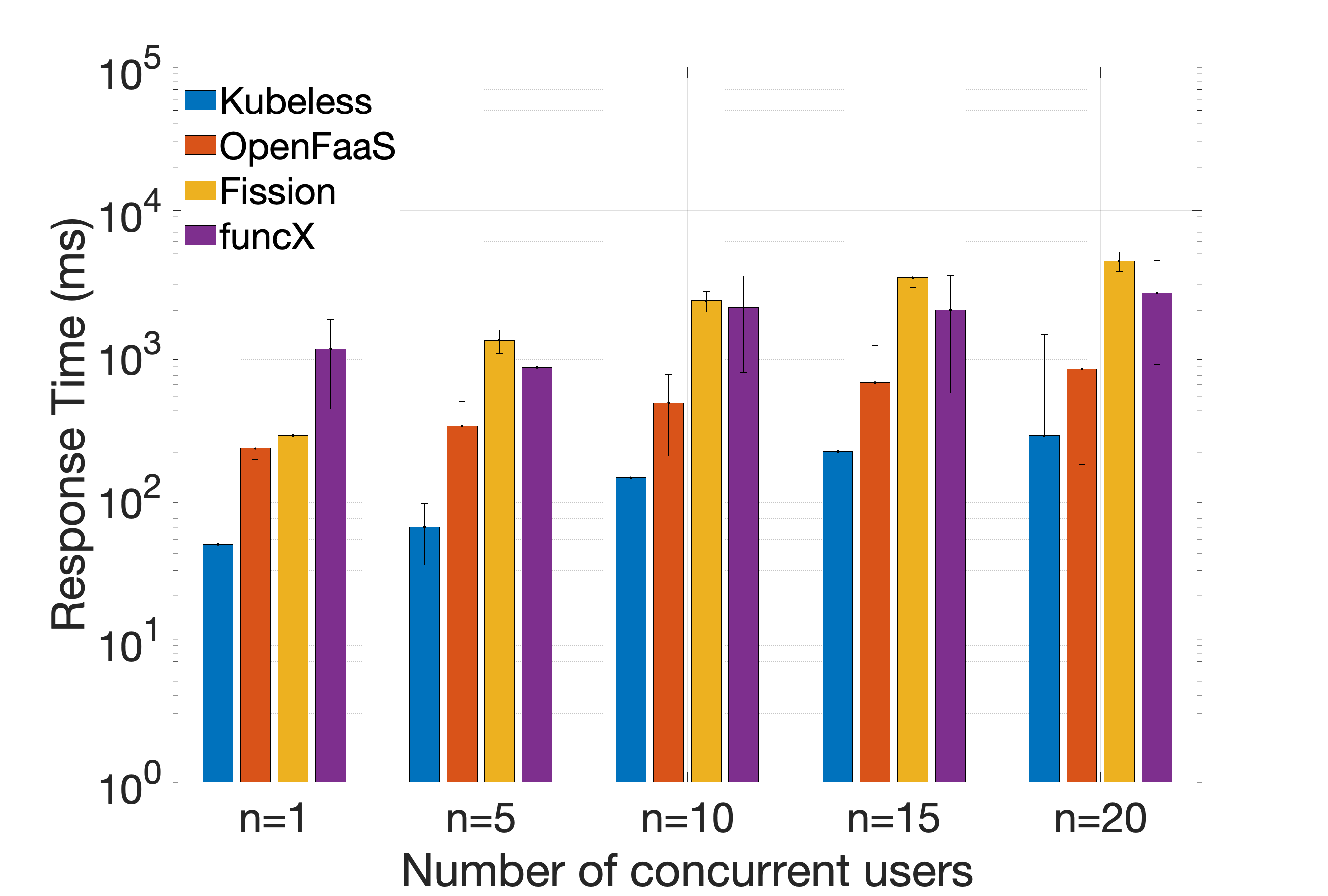}
         \caption{$Python$}
         \label{fig:y equals x}
     \end{subfigure}
     \hfill
     \begin{subfigure}[b]{0.45\textwidth}
         \centering
         \includegraphics[width=\textwidth]{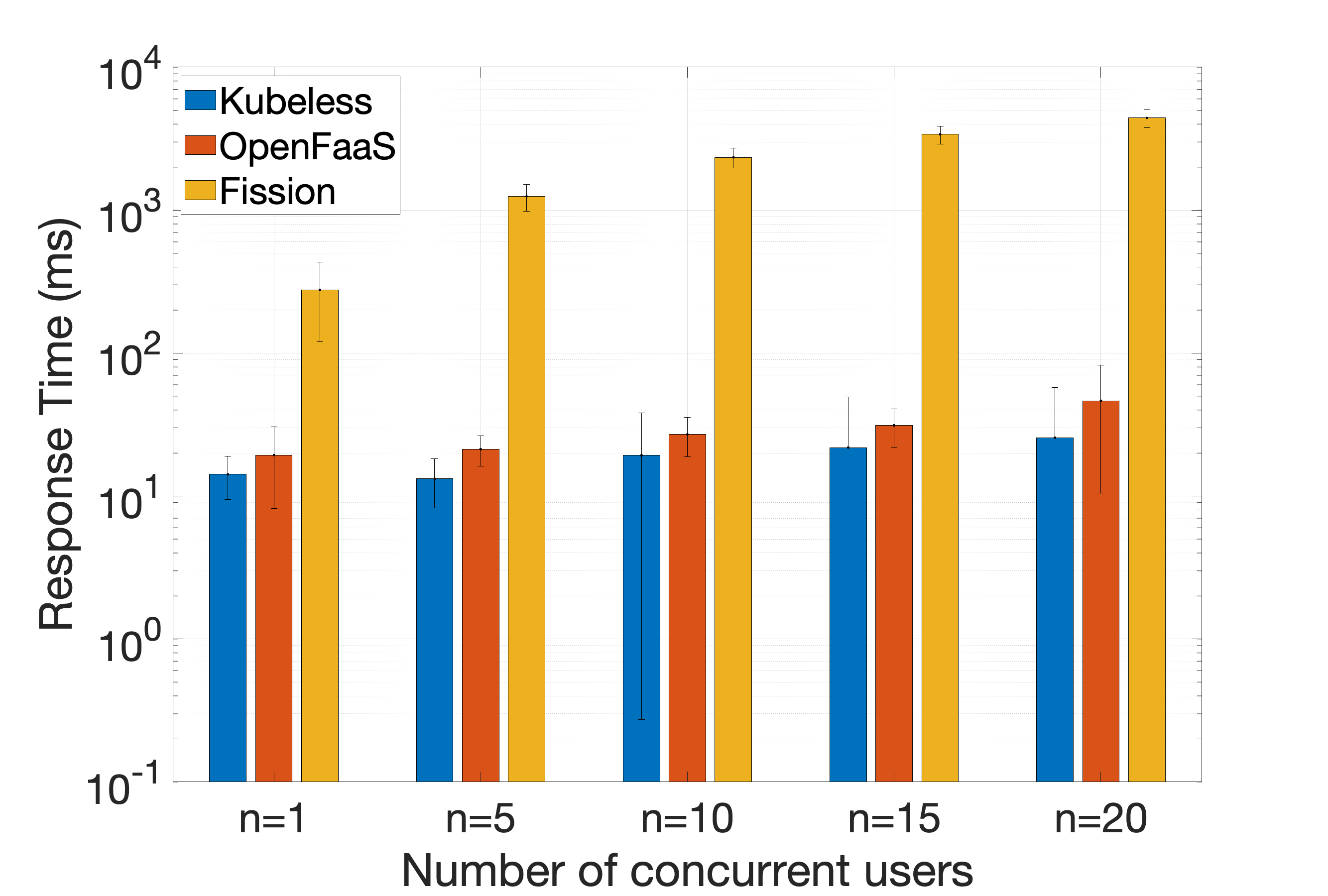}
         \caption{$Node.js$}
         \label{fig:three sin x}
     \end{subfigure}
  
        \caption{Average response time for Helloworld function for different frameworks in Python and Node.js}
        \label{fig:response_time}
\end{figure*}

The result of this experiments is plotted in Fig.~\ref{fig:response_time} which shows the average response time of each serverless framework for both Python and Node.js functions. 
The lowest average response time is achieved by Kubeless in all scenarios. We observe that Kubeless and OpenFaaS maintain a response time below 50 ms and 800 ms across all scenarios for Node.js and Python, respectively. We also note that the response times have steady increases with the number of concurrent users for Kubbeless and OpenFaas. However, for Fission and funcX, there is a higher increase in response time with the number of users which is related to the architecture of these frameworks to dispatch users' requests to available pods. We also observed that average response time for Node.js is considerably lower than Python function except with Fission which are about the same. The speed up ratio ($Python/Node.js$) is increased by the number of concurrent users and it reaches 10.3 for Kubeless for 20 concurrent users and 19.9 times for OpenFaaS for 15 concurrent users. 

\begin{figure*}
     \centering
     \begin{subfigure}[b]{0.45\textwidth}
         \centering
         \includegraphics[width=\textwidth]{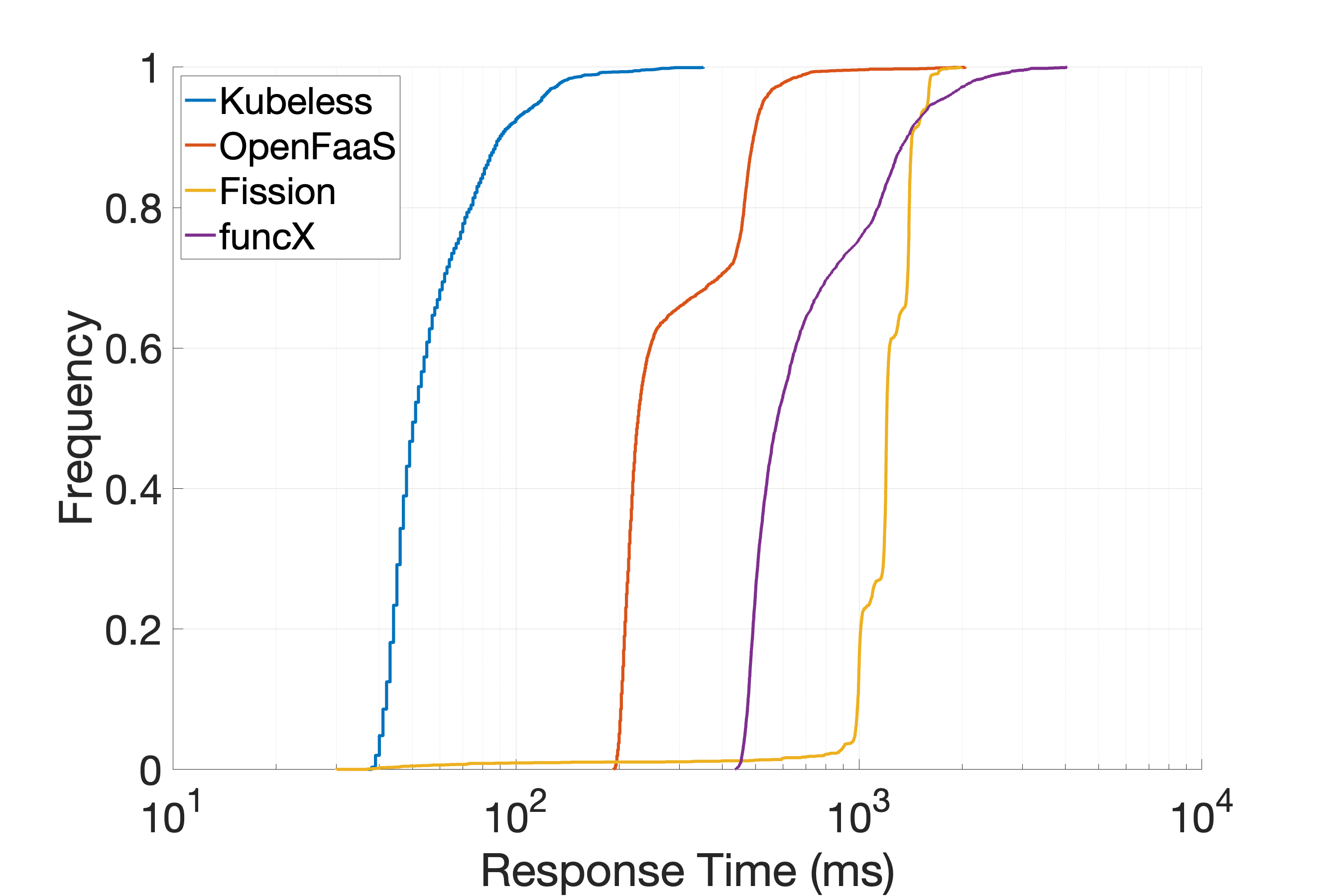}
         \caption{$Python$}
         \label{fig:y equals x}
     \end{subfigure}
     \hfill
     \begin{subfigure}[b]{0.45\textwidth}
         \centering
         \includegraphics[width=\textwidth]{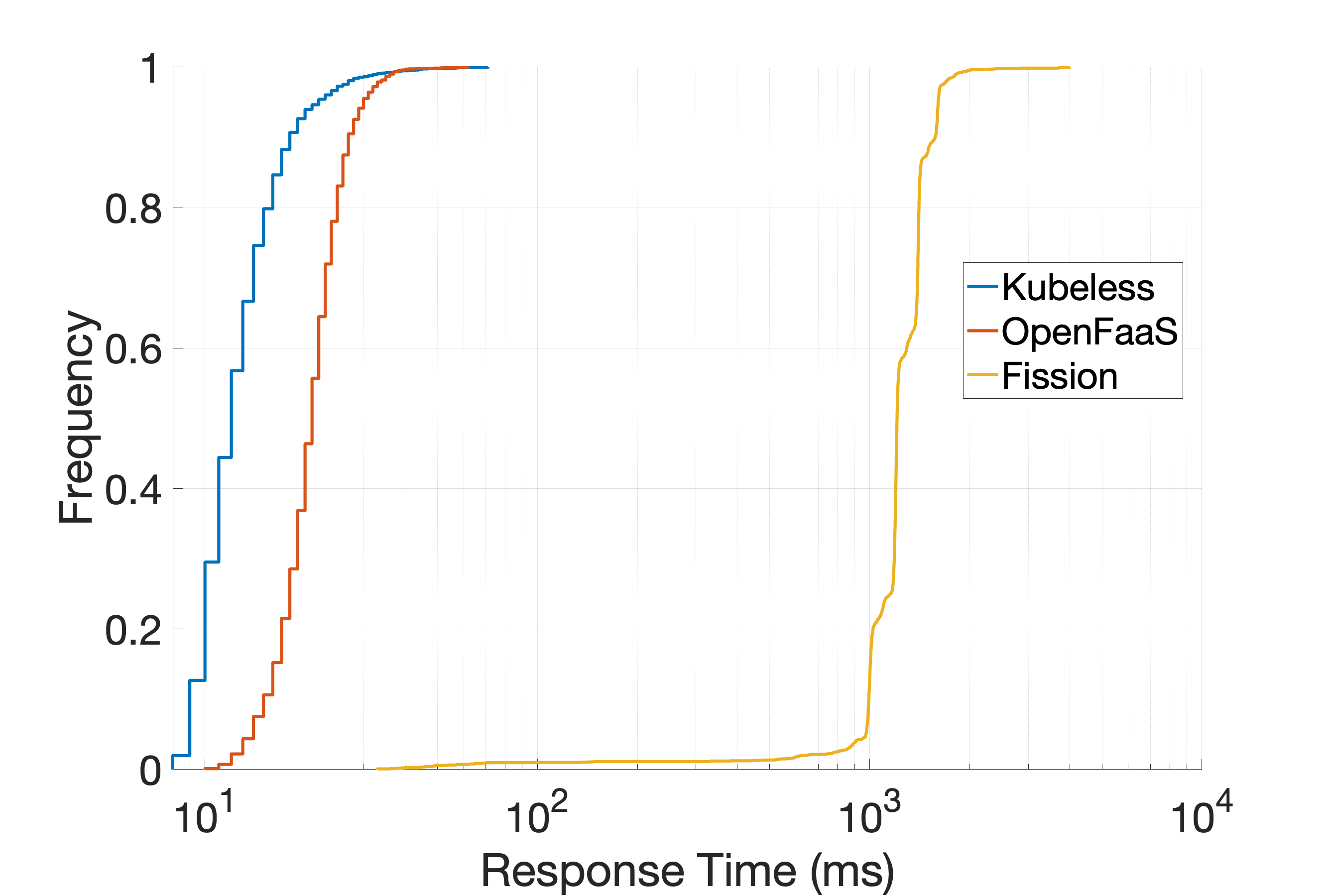}
         \caption{$Node.js$}
         \label{fig:three sin x}
     \end{subfigure}
  
        \caption{CDF of response time for Helloworld function ($n=5$) for different frameworks in Python and Node.js}
        \label{fig:cdf_5}
\end{figure*}

\begin{figure*}
     \centering
     \begin{subfigure}[b]{0.45\textwidth}
         \centering
         \includegraphics[width=\textwidth]{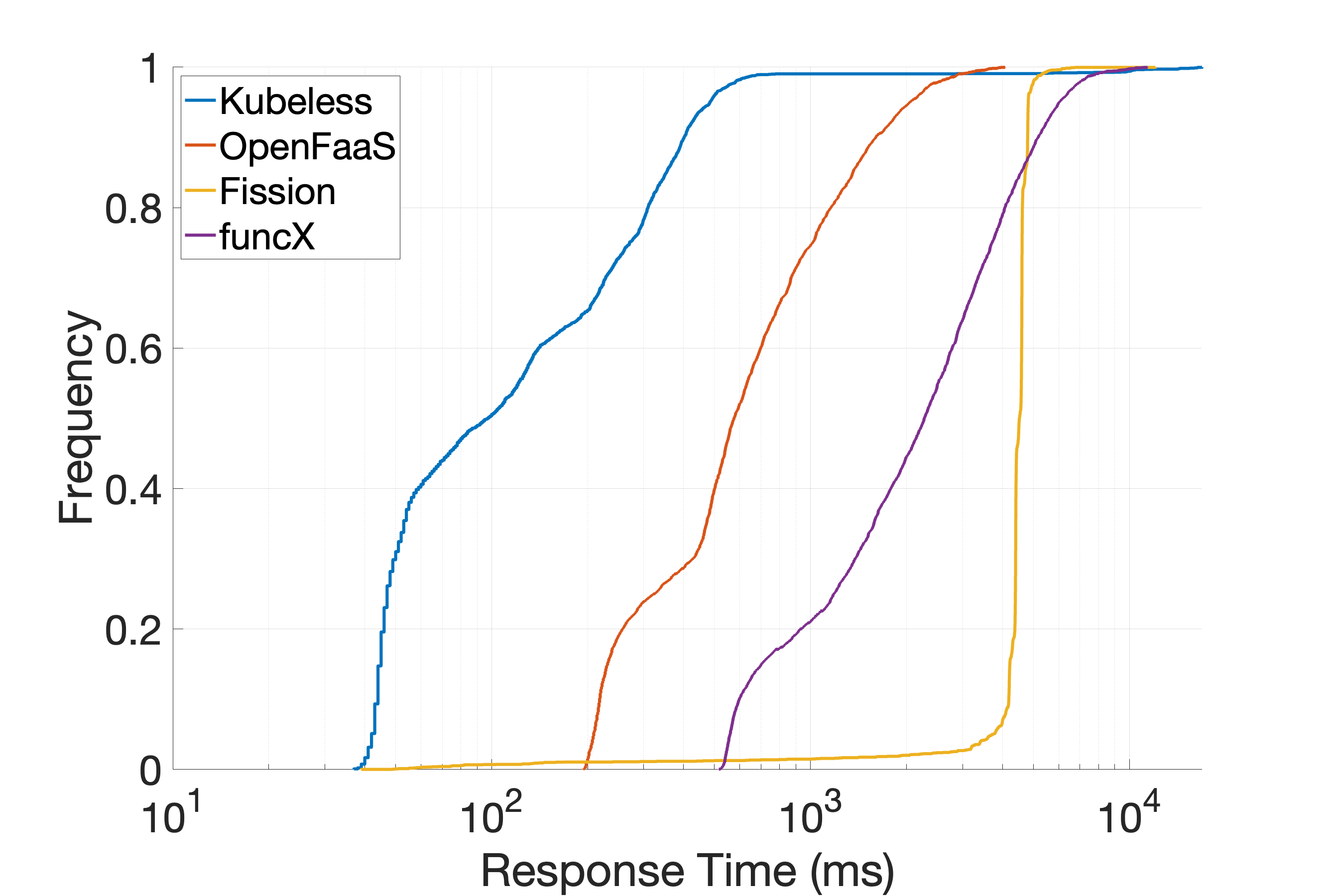}
         \caption{$Python$}
         \label{fig:y equals x}
     \end{subfigure}
     \hfill
     \begin{subfigure}[b]{0.45\textwidth}
         \centering
         \includegraphics[width=\textwidth]{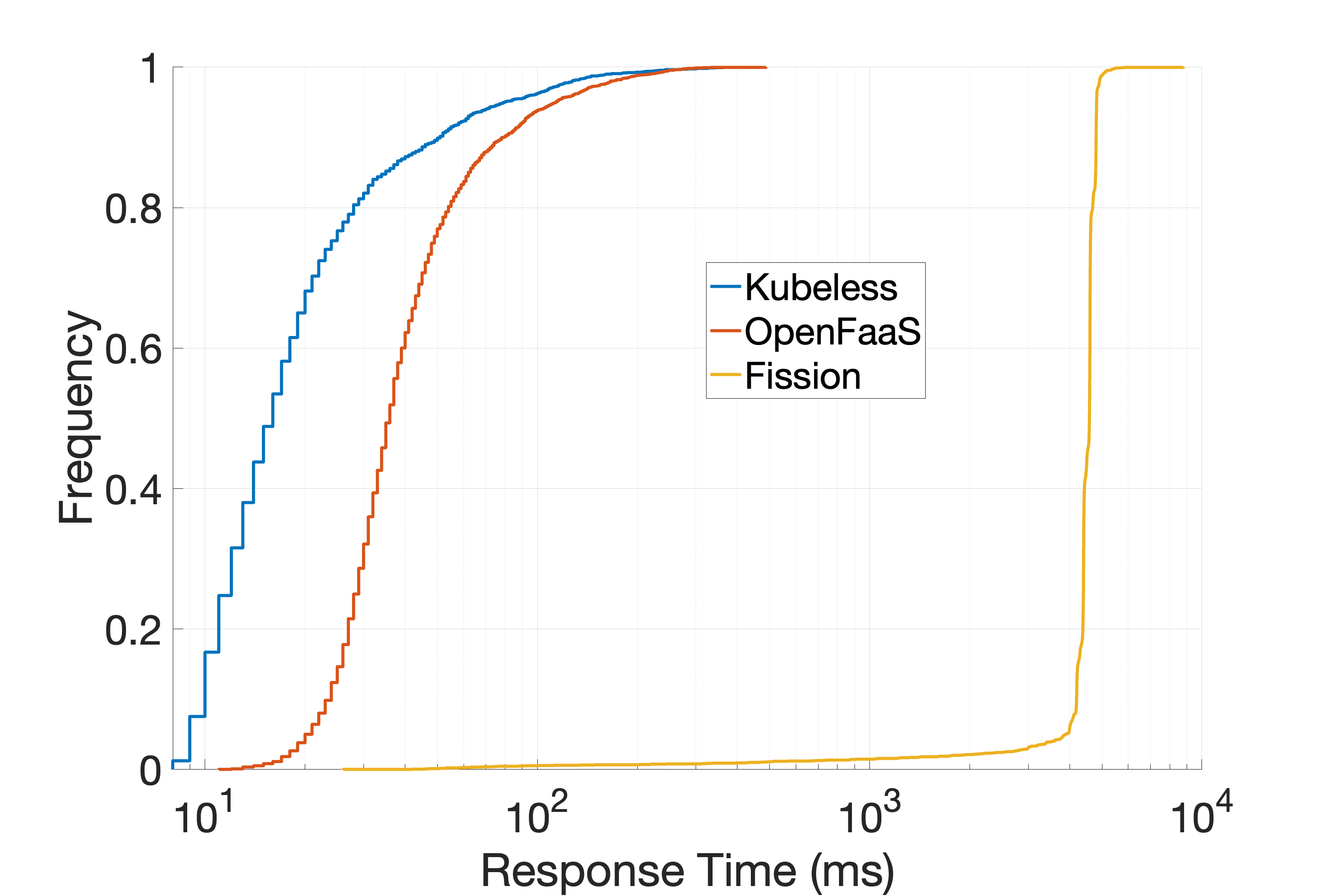}
         \caption{$Node.js$}
         \label{fig:three sin x}
     \end{subfigure}
  
        \caption{CDF of response time for Helloworld function ($n=20$) for different frameworks in Python and Node.js}
        \label{fig:cdf_20}
\end{figure*}

Fig.~\ref{fig:cdf_5} and Fig.~\ref{fig:cdf_20} show the cumulative distribution function of the response time (on log scale) for 5 and 20 concurrent users for all frameworks in both Python and Node.js, respectively. As can be seen, the system performance while running Node.js function is quite stable for both Kubeless and OpenFaaS which reveals the suitability of this language for latency-sensitive applications. For Python function, the system performance is less stable for the higher number of concurrent users with long-tail distributions for the response time as depicted in these figures. In addition, the performance gap between Kubeless and other frameworks are much bigger for the Python function in compare with the Node.js function. For both funcX and Fission, the response time is slightly higher than other frameworks which is increased by the number of concurrent users. We also noticed that for one user, the response time of Fission is comparable to OpenFaaS but this is not sustainable with increasing the number of concurrent users.

\begin{figure*}
     \centering
     \begin{subfigure}[b]{0.45\textwidth}
         \centering
         \includegraphics[width=\textwidth]{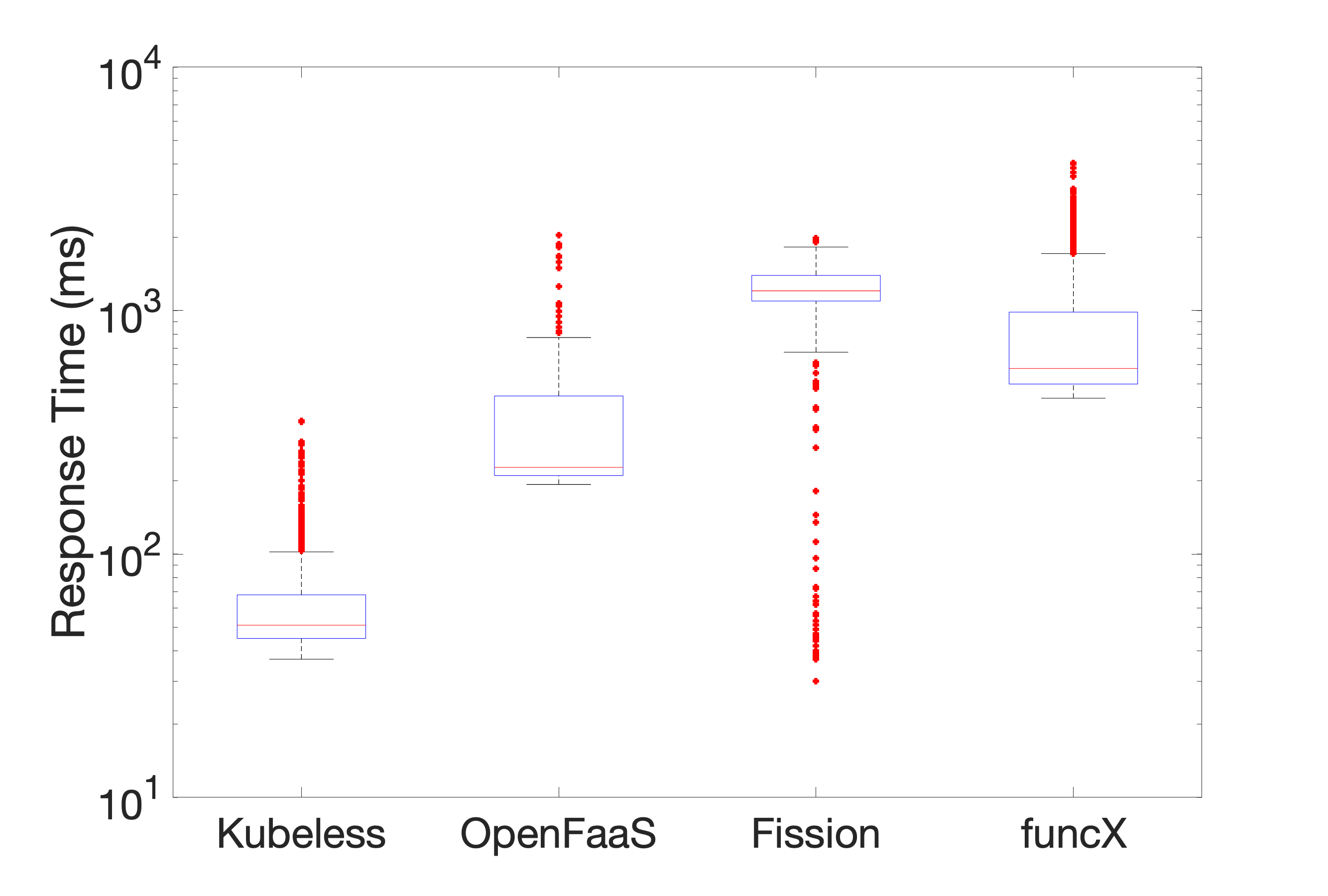}
         \caption{$Python$}
         \label{fig:y equals x}
     \end{subfigure}
     \hfill
     \begin{subfigure}[b]{0.45\textwidth}
         \centering
         \includegraphics[width=\textwidth]{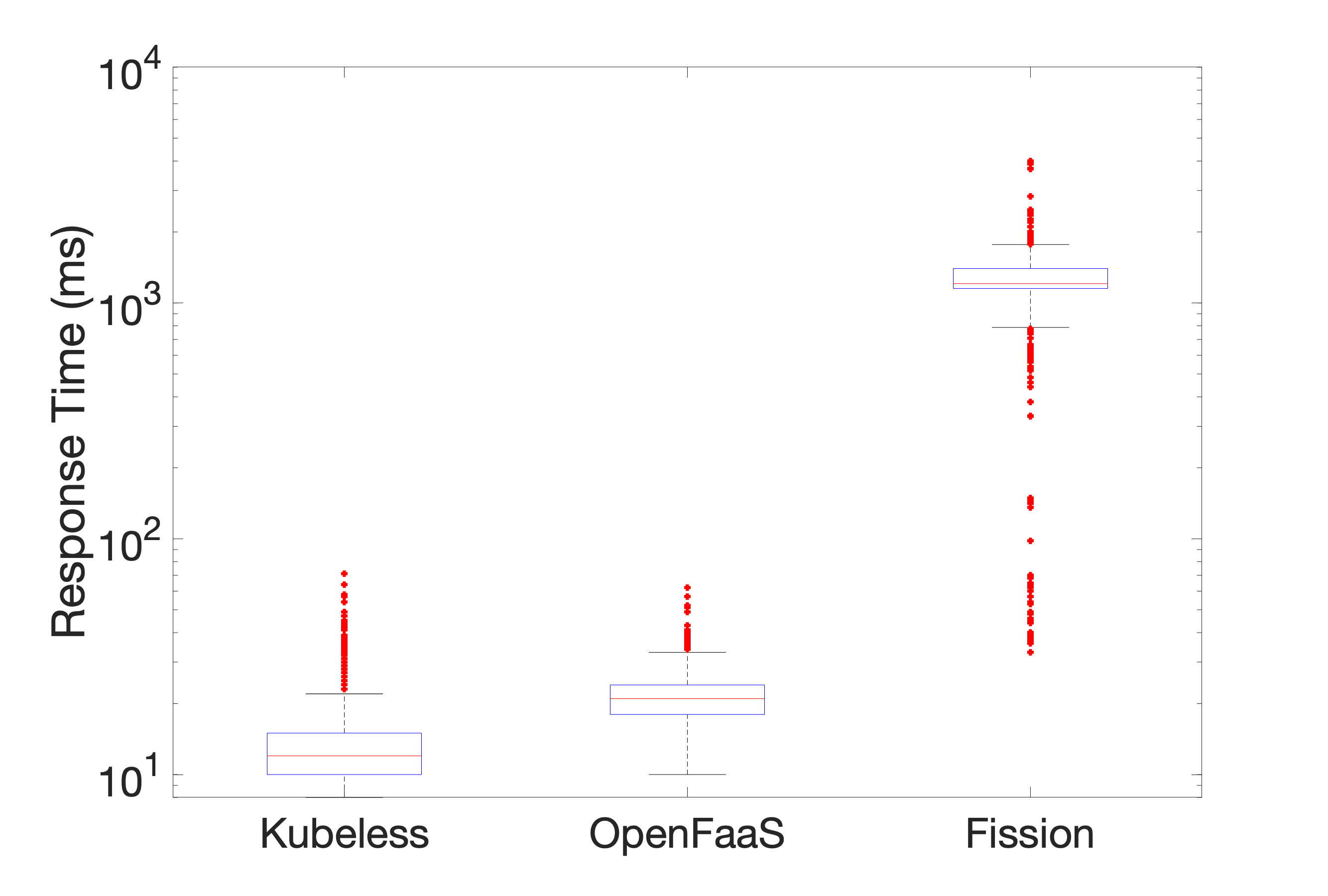}
         \caption{$Node.js$}
         \label{fig:three sin x}
     \end{subfigure}
  
        \caption{Box plots of response time for Helloworld function ($n=5$) for different frameworks in Python and Node.js}
        \label{fig:boxplot_5}
\end{figure*}

\begin{figure*}
     \centering
     \begin{subfigure}[b]{0.45\textwidth}
         \centering
         \includegraphics[width=\textwidth]{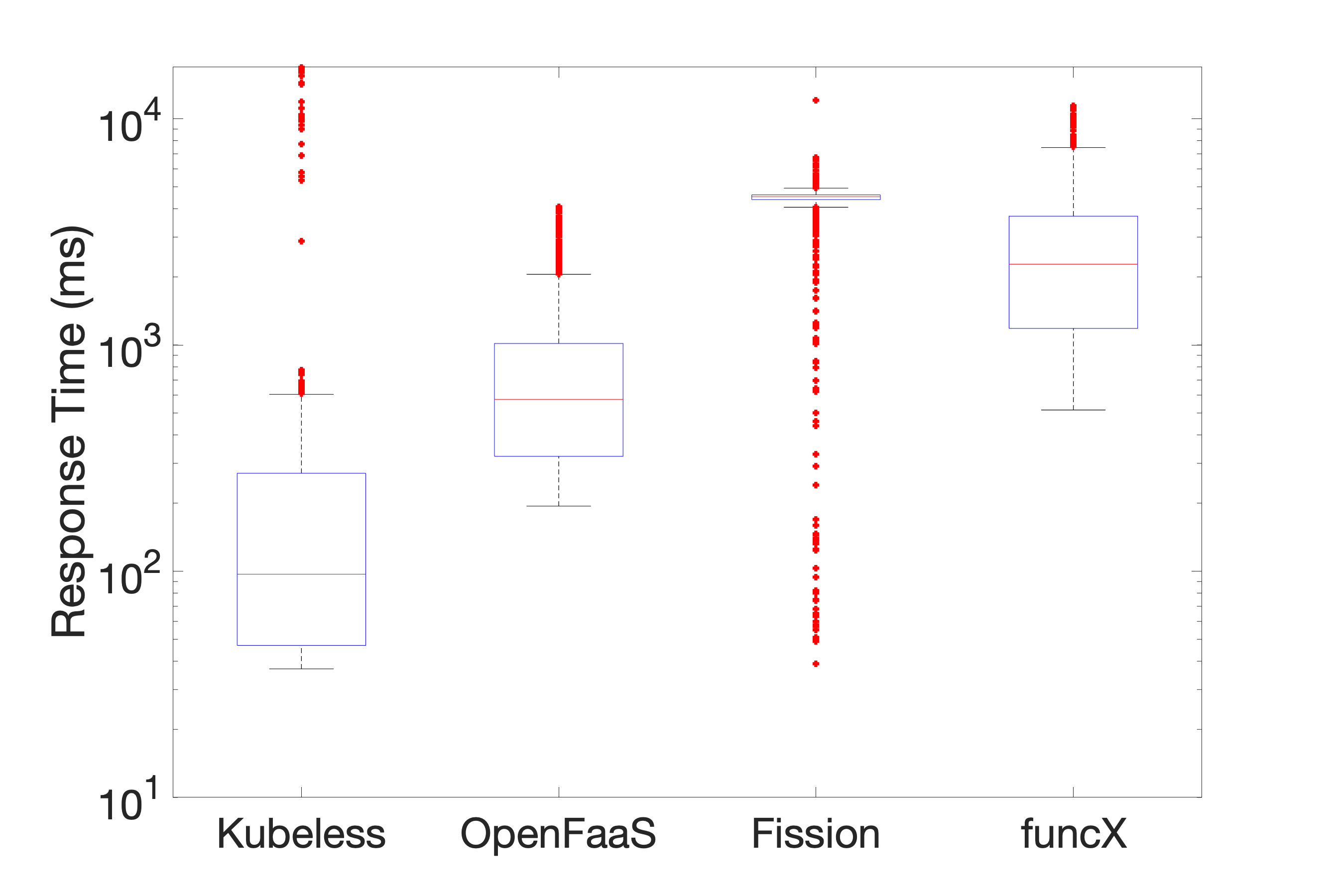}
         \caption{$Python$}
         \label{fig:y equals x}
     \end{subfigure}
     \hfill
     \begin{subfigure}[b]{0.45\textwidth}
         \centering
         \includegraphics[width=\textwidth]{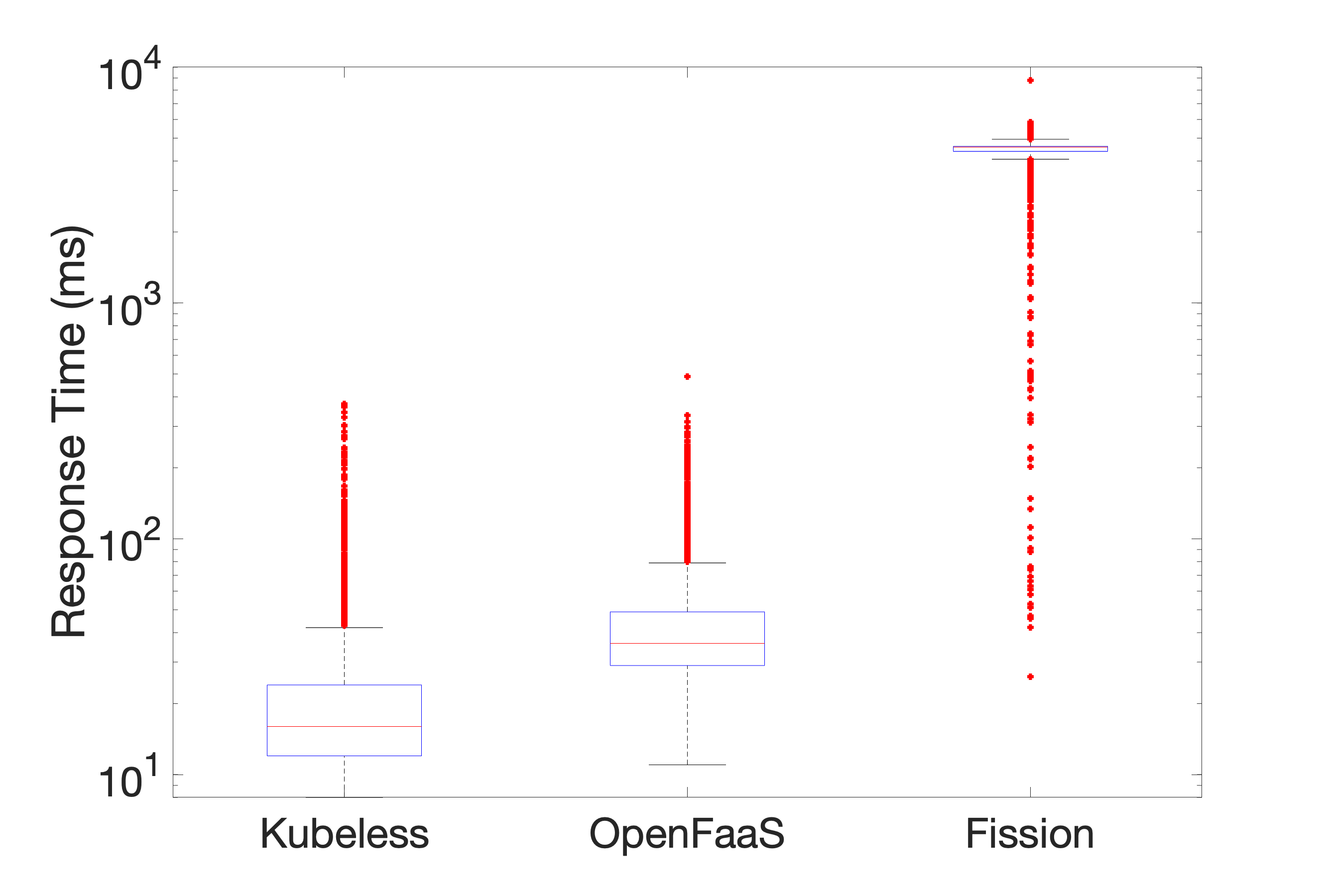}
         \caption{$Node.js$}
         \label{fig:three sin x}
     \end{subfigure}
  
        \caption{Box plots of response time for Helloworld function ($n=20$) for different frameworks in Python and Node.js}
        \label{fig:boxplot_20}
\end{figure*}

A closer examination of the results reveals that the response
times for Fission have a significant number of outliers as the
concurrency of requests increases more than one user (i.e., $n=1$). 
As an example, Fig.~\ref{fig:boxplot_5} and Fig.~\ref{fig:boxplot_20} show the response times (on log scale) for 5 and 20 concurrent users for all frameworks in both Python and Node.js, respectively. All frameworks have consistent response time except Fission which has several outliers in both functions. This performance degradation might be related to the \textit{router} component of Fission that forwards all incoming HTTP requests to the appropriate function. This component has a potential to become a bottleneck as the workload demand increases. For other frameworks such as Kubeless and OpenFaaS which are based on native Kubernetes components, there is more performance stability as they utilize the Kubernetes Ingress controller to route requests and balance the load.


\subsection{Impact of autoscaling}
In this section, we evaluate the impact of autoscaling on the response
time in all the serverless frameworks. For this experiment, we used both ML models as listed in Table~\ref{workload}. We choose to scale functions based on CPU utilization and set the threshold to 50\%, which is the trigger for creation of more function replicas in pods. For the autoscaling experiment, all frameworks start with one pod. For the experiment without autoscaling, the setup was one pod per edge device (i.e., 4 pods in total).
Please note that we did not set any resource request and limit for pods.
All frameworks, except funcX use the Kubernetes Horizontal Pod Autoscaler to perform scaling based on CPU utilization. We use the JMeter tool to send 1000 requests with 1 and 5 concurrent users and repeat each experiment 3 times. Given the resource limitations we could not execute more concurrency for all the frameworks.

\begin{figure*}
     \centering
     \begin{subfigure}[b]{0.45\textwidth}
         \centering
         \includegraphics[width=\textwidth]{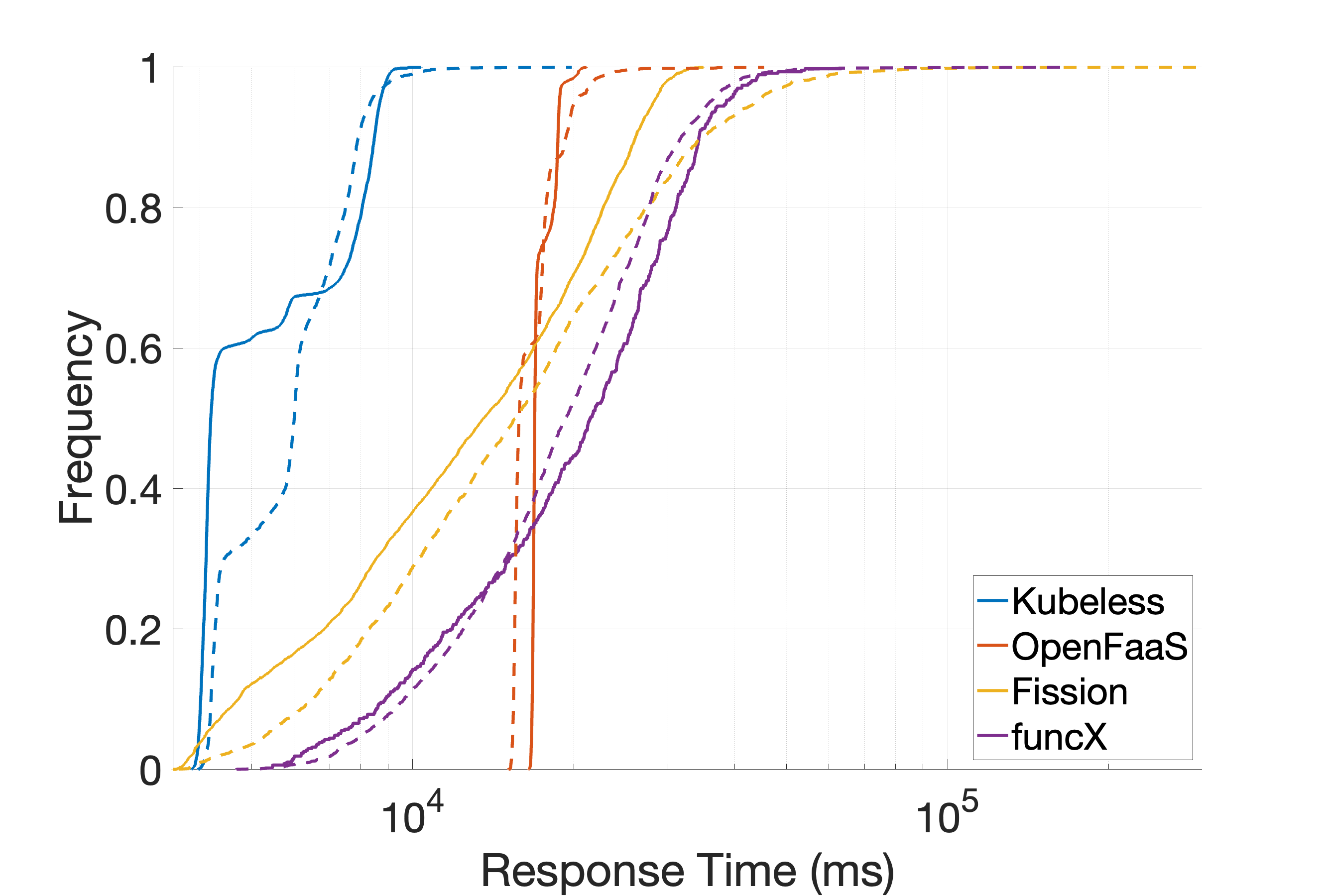}
         \caption{$n=1$}
         \label{fig:y equals x}
     \end{subfigure}
     \hfill
     \begin{subfigure}[b]{0.45\textwidth}
         \centering
         \includegraphics[width=\textwidth]{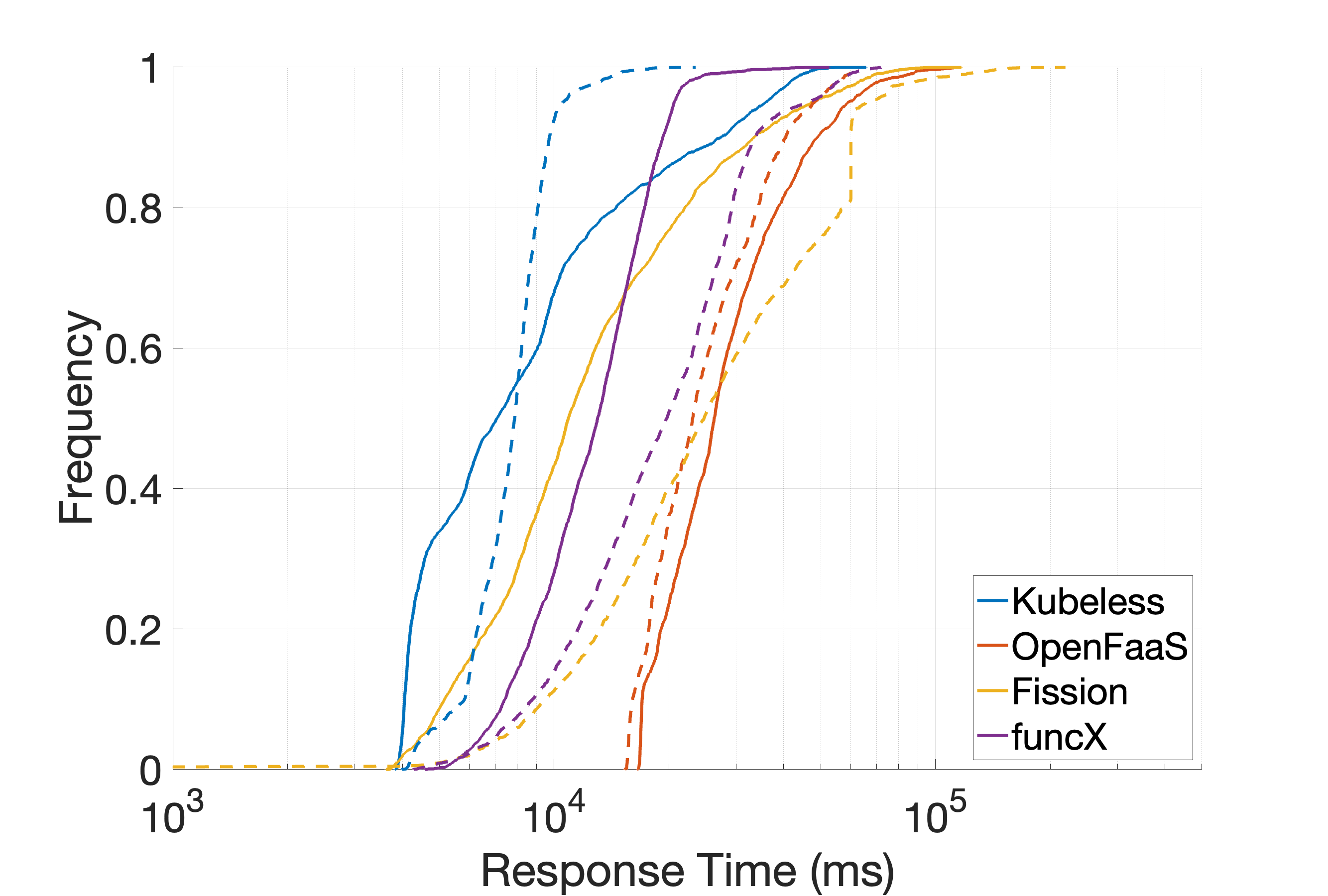}
         \caption{$n=5$}
         \label{fig:three sin x}
     \end{subfigure}
  
        \caption{CDF of response time for CNN function for different frameworks (solid lines show results with autoscaling, dashed lines show results without autoscaling)}
        \label{fig:cdf_cnn}
\end{figure*}

\begin{figure*}
     \centering
     \begin{subfigure}[b]{0.45\textwidth}
         \centering
         \includegraphics[width=\textwidth]{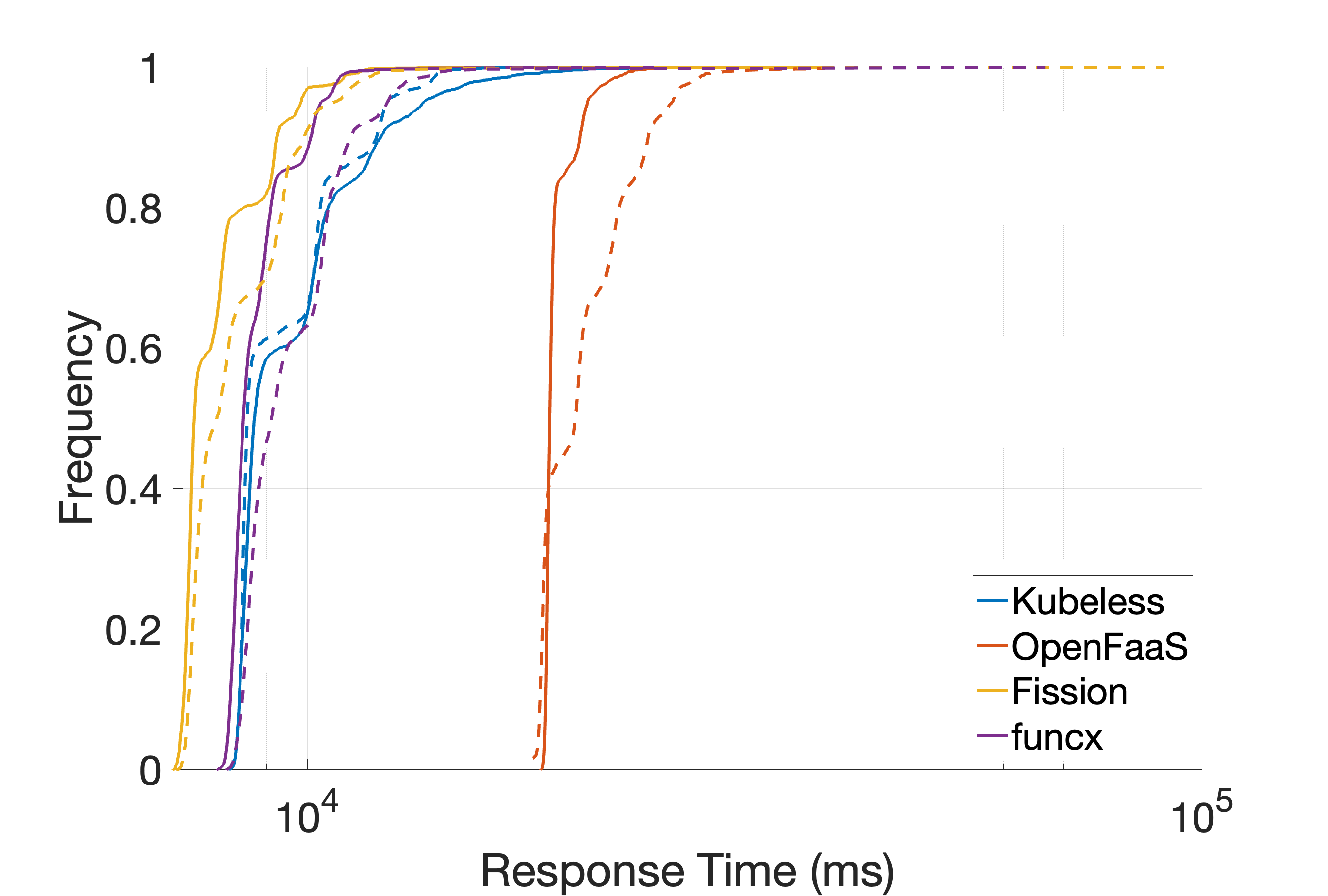}
         \caption{$n=1$}
         \label{fig:y equals x}
     \end{subfigure}
     \hfill
     \begin{subfigure}[b]{0.45\textwidth}
         \centering
         \includegraphics[width=\textwidth]{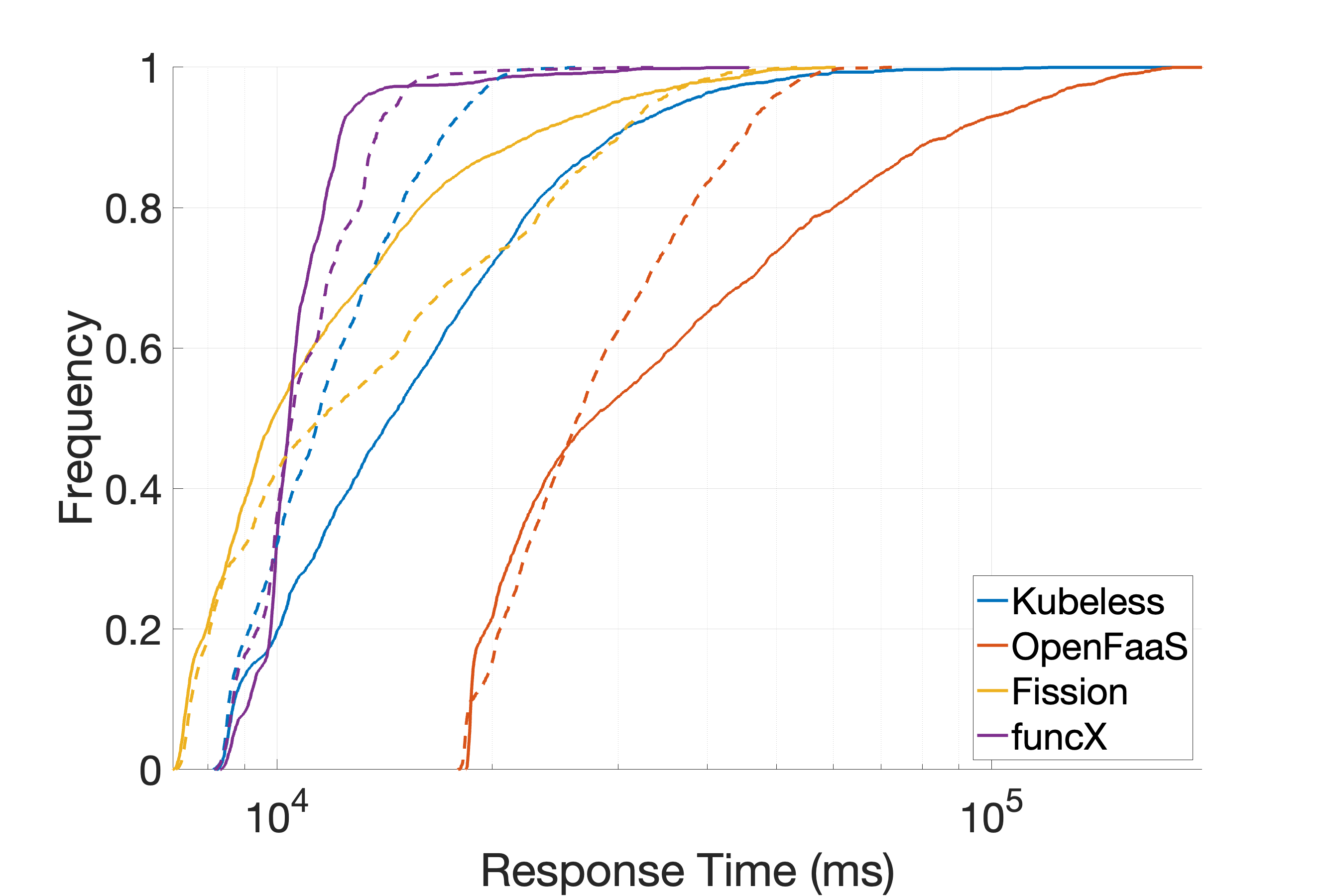}
         \caption{$n=5$}
         \label{fig:three sin x}
     \end{subfigure}
  
        \caption{CDF of response time for LSTM function for different frameworks (solid lines show results with autoscaling, dashed lines show results without autoscaling)}
        \label{fig:cdf_lstm}
\end{figure*}

Fig.~\ref{fig:cdf_cnn} and Fig.~\ref{fig:cdf_lstm} show the results for experiments for all the frameworks for 1 and 5 concurrent users with autoscaling for CNN and LSTM functions, respectively. In these figures, solid lines represent results of autoscaling and dashed lines show the same configuration without autoscaling while colors differentiate between different frameworks.
These results show that for CNN function, Kubeless achieved the best response time with and without autoscaling. However, Fission is the only framework that can leverage the autoscaling to improve the response time by 27\% and 100\% for $n=1$ and $n=5$, respectively. For LSTM function, Fission and funcX can provide a better response time which is due to their ability for more efficient autoscaling as presented in Fig.~\ref{fig:cdf_lstm}. Fission can leverage the autoscaling to improve the response time by 5\% and 30\% for $n=1$ and $n=5$, respectively.

\begin{figure*}
     \centering
     \begin{subfigure}[b]{0.45\textwidth}
         \centering
         \includegraphics[width=\textwidth]{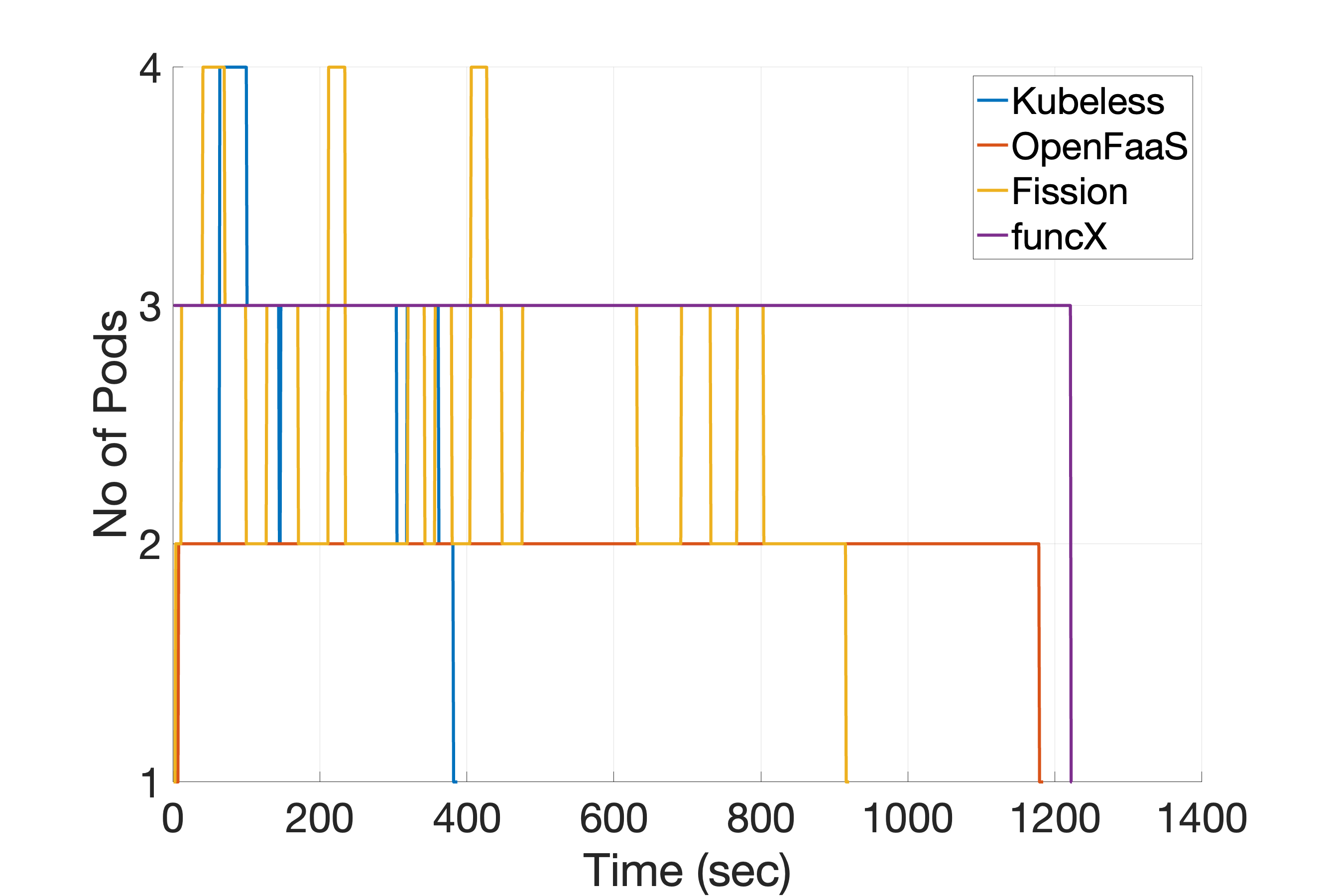}
         \caption{$n=1$}
         \label{fig:y equals x}
     \end{subfigure}
     \hfill
     \begin{subfigure}[b]{0.45\textwidth}
         \centering
         \includegraphics[width=\textwidth]{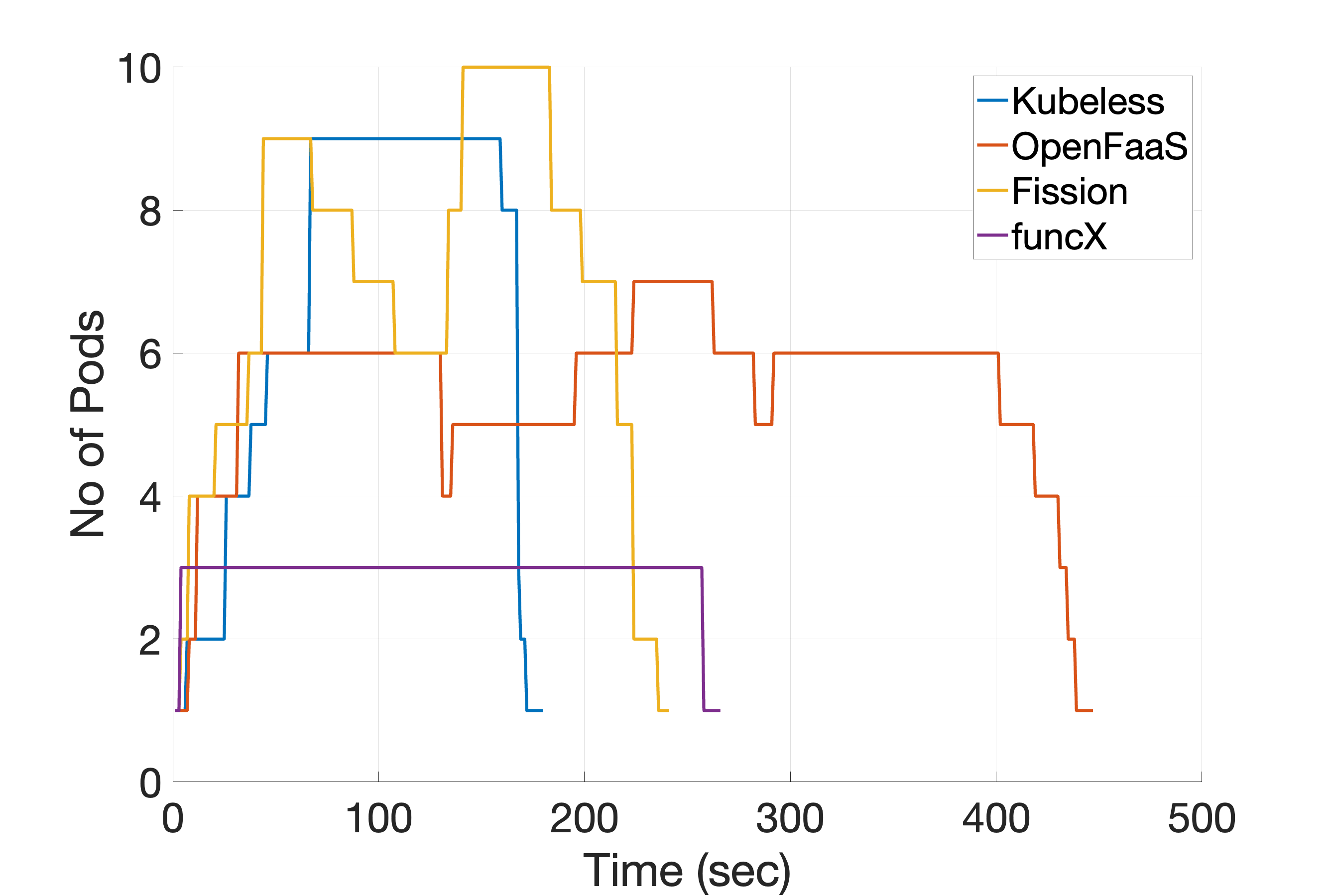}
         \caption{$n=5$}
         \label{fig:three sin x}
     \end{subfigure}
  
        \caption{Number of pods for autoscaling of CNN function for different frameworks}
        \label{fig:pod_cnn}
\end{figure*}

\begin{figure*}
     \centering
     \begin{subfigure}[b]{0.45\textwidth}
         \centering
         \includegraphics[width=\textwidth]{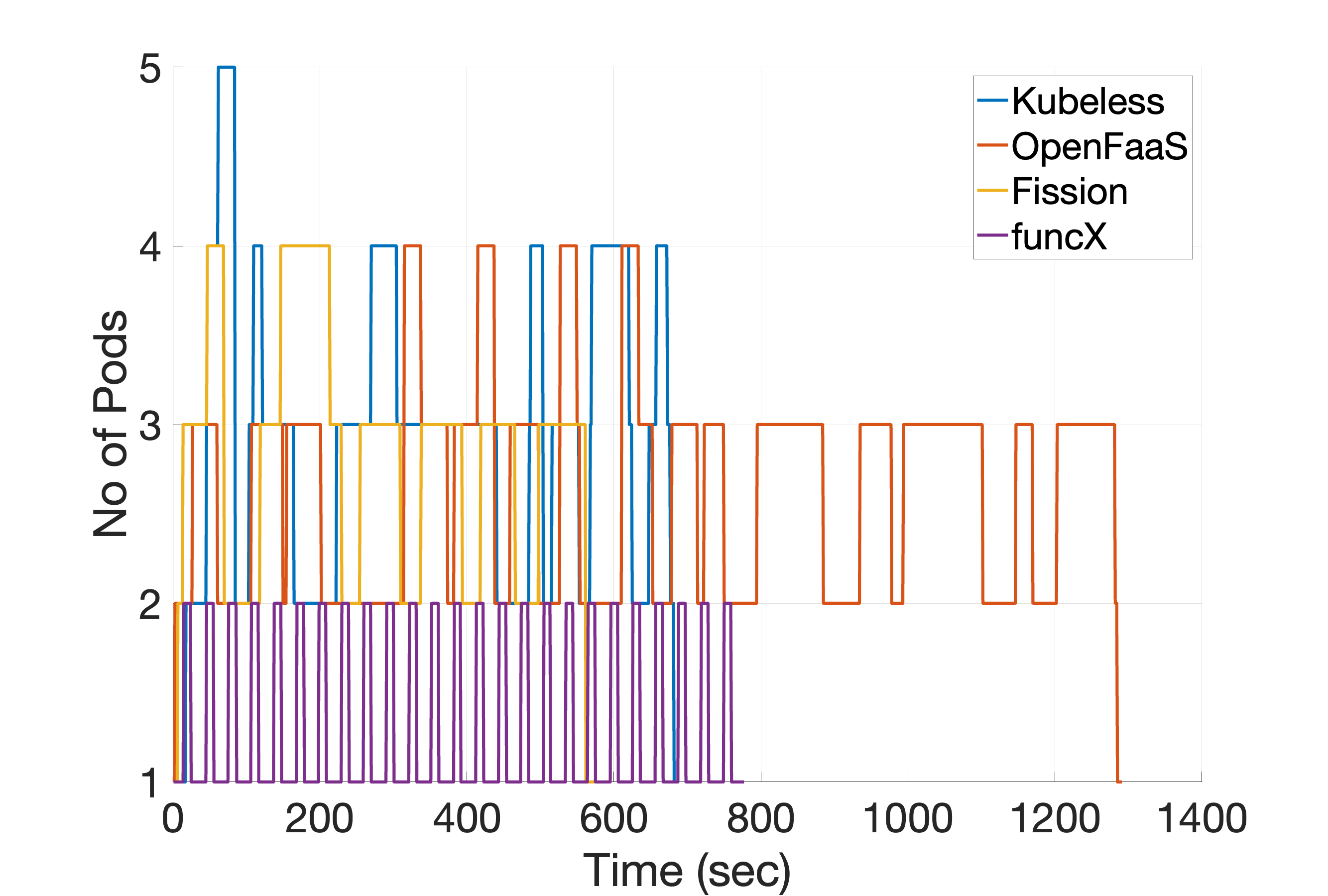}
         \caption{$n=1$}
         \label{fig:y equals x}
     \end{subfigure}
     \hfill
     \begin{subfigure}[b]{0.45\textwidth}
         \centering
         \includegraphics[width=\textwidth]{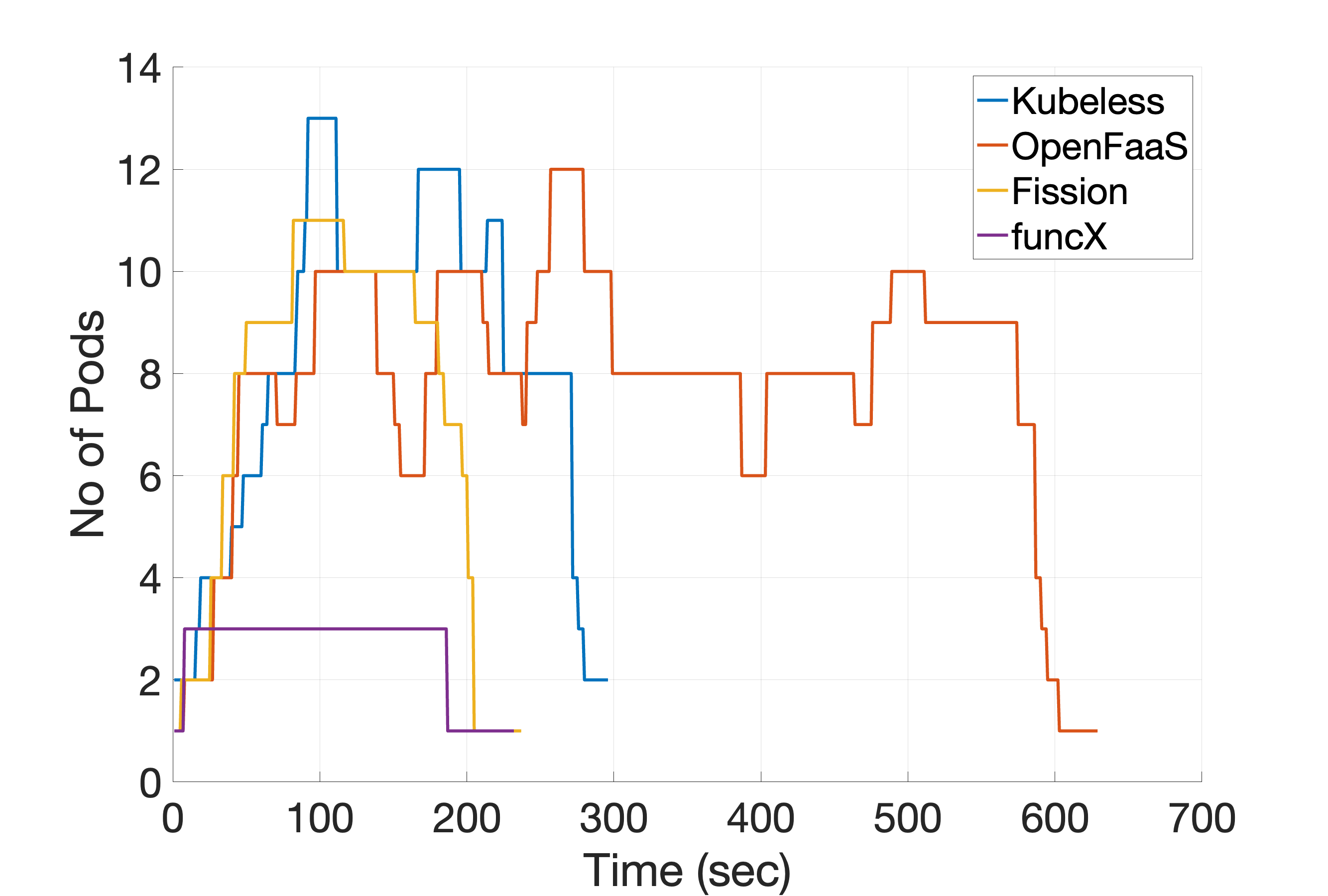}
         \caption{$n=5$}
         \label{fig:three sin x}
     \end{subfigure}
  
        \caption{Number of pods for autoscaling of LSTM function for different frameworks}
        \label{fig:pod_lstm}
\end{figure*}

We also examine the variation of pod numbers during a single iteration of the experiment for both CNN and LSTM functions as plotted in Fig.~\ref{fig:pod_cnn} and Fig.~\ref{fig:pod_lstm}. As it can be seen in these figures,  Fission and Kubeless are adding more pods in response to an increase in the CPU utilization due to higher workload. OpenFaaS handle autoscaling with a more conservative approach and add less pods compared to the other two frameworks. As mentioned earlier, funcX does not support autoscaling with CPU utilization, so there is not much change in the number of pods. We also notice that the total duration of the experiment is longer for OpenFaaS due to the higher average response time where all 1000 requests need to be processed. This is justified by the fact that each JMeter's thread is executed in a synchronized fashion and waits for a response before sending the next request.

In addition, we examine the ratio of successfully received responses (i.e., success rate) under different levels of concurrent requests for both machine learning workloads. For one user, both Kubeless and Fission show a reliable behaviour with 100\% success rate, but OpenFaaS and funcX have 1-2\% of error. This pattern is repeated for higher concurrency levels, but we observed less success rate for funcX in this case with an error up to 3\%.

\subsection{Model training}
In this section, we consider the performance of edge platforms for machine learning training. As stated in section~\ref{setup}, for the machine learning workloads we trained the models on a x86 machine and execute the model inference as the serverless function. As a new experiment, we run the model training on a Raspberry Pi with an ARM processor and reported the results in Table~\ref{training}. It is noted that, the CNN model has a higher accuracy, but it takes a longer amount of time to train over both processor architectures. However, shorter run times are observed for CNN classification on the edge compared to LSTM as illustrated in Fig.~\ref{fig:cdf_cnn} and Fig.~\ref{fig:cdf_lstm}. A slower execution time for classification observed for the LSTM model in part relates to its higher cell structure complexity and its sequential execution model, which is less amenable to parallel processing compared to a CNN network architecture. 

Although the CNN model is clearly a more preferable choice in terms of accuracy, implementing two different deep learning network architectures highlights their significantly different behaviours, such as during pod recruitment in autoscaling, that may favour one architecture over another. These algorithmic considerations become very relevant during edge processing conditions, where resource limitations exacerbate small differences in low level container behaviour, in comparison to a less constrained CPU processing environment. The choice of edge-based machine learning models needs to be carefully considered for both classification and training modes, but more so in the latter case due to the significantly heavier computational demands involved. As derived from Table~\ref{training}, the $ARM/x86$  training time ratio of 7.38 for LSTM compared to 15.58 for CNN demonstrates the added processing overhead that can potentially be saved by simply choosing a more suitable machine learning model for edge training. A compromise may need to be made between choosing an efficient model that is optimised for training on the edge, despite resulting in an overall decreased classification accuracy. 

\begin{table}[htbp]
\caption{Machine learning training results}
\centering
\label{training}
\begin{tabular}{|l|c|c|c|}
\hline
\multirow{2}{*}{\textbf{Machine learning model}}  &  \multicolumn{2}{l|}{\textbf{Training time (sec)}}  & \multirow{2}{*}{\textbf{Accuracy}}\\
\cline{2-3}
    & \textbf{x86} & \textbf{ARM} &\\
\hline
CNN &   435 & 6780 & 0.99\\
\hline
LSTM & 163 & 1203 & 0.90\\
\hline
\end{tabular}
\end{table}

\subsection{Discussions}
Our experimental results demonstrate that Kubeless has the most
consistent performance across different scenarios for edge computing. We observed that this framework can provide a latency of 10-50 ms for up to 20 concurrent users which is satisfactory for a wide range of latency sensitive applications. The main reason is the simple architecture of Kubeless and the use of native Kubernetes services which has a low overhead for an edge computing platform. VMWare has decided to stop driving and updating Kubeless, however since this is an open source framework, the research community is able to use, modify and improve it for edge computing platforms.

In other frameworks such as OpenFaaS and Fission, there are internal components such as \textit{watchdog} or \textit{router} that have the potential to be a bottleneck in some of the experiments with high level of concurrency which lead to performance degradation for edge computing platforms. We examined the supported auto-scaling algorithms and observed that the performance of Fission and in some cases Kubeless increases with the number of pods, while OpenFaaS showed the same performance regardless to the increased number of pods.

funcX is a relatively new serverless framework with some interesting features including independency from Kubernetes with a centralized cloud-based function registration. While the reported latency is higher than other frameworks due to extra communication overhead with the cloud, the scalability of this framework for machine learning workloads for higher number of concurrent users is close to or better than other frameworks. 

\section{Conclusions and Future Work}
\label{conclusion}

In this paper, the performance evaluation of serverless edge computing under different machine learning workloads has been analyzed. The selected test-bed is based on the most popular serverless frameworks with distinct features, including Kubeless, OpenFaaS, Fission and funcX. We found that Kubeless outperforms the other frameworks in terms of response time for basic workloads, and Fission has the worst performance for handling concurrent users. In addition, Node.js is a considerably faster language compared to Python and can be considered to be the selected deployment language for real-time functions. For machine learning workloads, we demonstrated that based on the structure and complexity of the model, frameworks such as Fission could have a better performance with limited concurrent users.  Moreover, if we do not set any resource limit on pods, it might be better not to use autoscaling in some cases as it has some performance degradation for frameworks such as OpenFaaS.
In future work, we intend to work on modifying some of the critical components of serverless frameworks, including the function handling and autoscaling modules, to improve their performance for edge computing platforms. 

\balance

\bibliographystyle{IEEEtran}
\bibliography{references}
%




\end{document}